\documentclass[final,5p,times,twocolumn]{elsarticle}

\usepackage{float}

\usepackage[all]{nowidow}
\usepackage{csquotes}

\usepackage{mathtools}

\usepackage[detect-all]{siunitx}
\usepackage{dcolumn}
\usepackage{booktabs}
\usepackage{color}
\biboptions{sort&compress}
\usepackage[all]{nowidow}
\usepackage{amsmath}
\usepackage{amssymb}

\journal{Nuclear Physics B}

\begin{document}

\begin{frontmatter}



%\title{Predictions for charged-hadron production in $p$-O collisions at LHC energies}

\title{In-medium $\Upsilon(1S,2S,3S)$ suppression in Pb-Pb collisions at $\sqrt{s_{NN}}=5.02$ TeV }


%\author{L.~Konrad$^{1}$, P.~Schulz$^{2}$ and G.~Wolschin$^{1}$}
\author{J.~Majonica and G.~Wolschin$^{1}$}
%\affiliation{organization={Institute for Theoretical Physics, Heidelberg University},%Department and Organization
%            addressline={Philosophenweg 16}, 
%            city={Heidelberg},
 %           postcode={69120}, 
  %          state={Baden Wurttemberg},
%            country={Germany}}
\noindent
{{
$^1$Institute for Theoretical Physics, Heidelberg University}\\%Department and Organization
\hspace{.6cm}        {Philosophenweg 16}\\
      {Heidelberg~
          {69120}\\ 
         {Baden W\"urttemberg},
          {Germany}}\\
          wolschin@uni-heidelberg.de\\\\

\begin{abstract}
We present model calculations for the in-medium suppression of the $\Upsilon(1S,2S,3S)$ states in $\sqrt{s_{NN}}=5.02$ TeV Pb-Pb collisions at the Large Hadron Collider in comparison with recent CMS data for all three spin-triplet $s$-wave states. The model parameters initial central temperature, and formation times for the $\Upsilon(nS)$ states
%, as well as formation times for the six $\Upsilon(nS)$ and $\chi_b(nP)$ states 
are determined in simultaneous $\chi^2$ minimizations with respect to the data, such that the sequential centrality- and transverse-momentum-dependent suppression of the observed states is reproduced.
\end{abstract}

%\includegraphics[scale=0.7]{fig4.eps}

%\item Nonlinear boson diffusion equation applied to gluon thermalization
%\item Analytical calculation of the time evolution
%\item Application to heavy-ion collisions at the LHC

}

\end{frontmatter}


\section{Introduction}
\label{intro}
The in-medium modification of quarkonia states such as $J/\psi$  {\cite{phe07,alice12i,alice20jhep,ms86}}  and $\Upsilon(nS)$ \cite{CMS-2012,cms17,star23} in relativistic heavy-ion collisions has turned out to be a sensitive indicator for quark-gluon plasma (QGP) properties. 
In central $\sqrt{s_{NN}}=5.02$ TeV Pb-Pb collisions, the relatively abundant charm-quarks produce interesting recombination effects following $J/\psi$-dissociation in the hot medium  \cite{alice17i,alice24}. Bottom-quarks, however, rarely recombine once the $\Upsilon(nS)$ bound states that emerge at short formation times have been dissociated in the hot medium of gluons and light quarks. Indeed, no indications for recombination effects in the nuclear suppression factor such as an extended flat or even increasing region for mid-central and central collisions are observed \cite{cms19}. 

In this work, we concentrate on bottomonium spectroscopy in the hot medium. We use
%take advantage of 
recent detailed CMS measurements \cite{cms24} of the $\Upsilon(1S,2S,3S)$ suppression in $\sqrt{s_{NN}}=5.02$ TeV Pb-Pb collisions at the Large Hadron Collider (LHC) to optimize the initial central temperature $T_0$ and formation times $\tau_\mathrm{F}^{nS}$ for the three observed states in simultaneous $\chi^2$ minimizations. These are the parameters in the model for $\Upsilon$ suppression in the hot quark-gluon medium that we have developed earlier, and used for predictions \cite{hnw17,gw20i} in $\sqrt{s_{NN}}=5.02$ TeV Pb-Pb \cite{cms24} and $\sqrt{s_{NN}}=200$ GeV Au-Au collisions \cite{star23}.

Whereas the heavy-quark formation times from initial hard partonic interactions in the very first stages of a relativistic collision can be estimated from the time-energy-uncertainty relation as $\tau_c\simeq 1/(4m_c)\simeq 0.04$ fm for $c$-quark pairs and $\tau_b\simeq 0.01$ fm for $b$-quark pairs, the subsequent in-medium formation time $\tau_\mathrm{F}$ of the corresponding quark-antiquark bound states is more difficult to determine. It depends on the time-dependent temperature $T(t)$ of the surrounding medium. For short formation times,
% -- as for the $\Upsilon(nS)$ states that we investigate here --,
the corresponding quarkonia states have a low probability to survive the in-medium processes such as screening, damping and gluodissociation, and their yield compared to 
the scaled vacuum production will be suppressed. Longer formation times correspond to higher survival probabilities because the time spent in the hot medium is shorter. 

In line with common expectation, early formation-time studies using in-medium quarkonia properties \cite{ko15} and more recent investigations such as \cite{ch24} have discussed formation times being inversely proportional to the binding energies with respect to the open-charm or open-bottom thresholds. However, the experimentally observed sequential suppression of the $\Upsilon(nS)$ states in Pb-Pb collisions at energies reached at LHC \cite{CMS-2012,cms17,cms19,alice19b,cms24} and also in Au-Au collisions at the Relativistic Heavy Ion Collider (RHIC) with $\sqrt{s_{NN}}=200$ GeV \cite{star23} could imply that the higher-lying states have shorter formation times, therefore remain longer in the hot quark-gluon medium and thus, are more strongly suppressed. Indeed, calculations indicate that it is difficult within our model to produce the sequential suppression seen in the data using formation times that are inversely proportional to the binding energies. Hence, the observed strong sequential suppression of $\Upsilon(2S,3S)$ might even imply shorter formation times compared to $\Upsilon(1S)$. 

It is recognized that the sequential $\Upsilon(nS)$ suppression can also be generated for state-independent formation times (or even for $\tau_\mathrm{F}\propto1/E_\mathrm{B}$) if the thermal widths of the excited states were substantially larger than the width of the ground state, as is the case in lattice studies of static quarkonia \cite{ch24}. 
%This is, however, not the case in our model calculation which -- 
%Different from lattice calculations, 
Our model \cite{ngw13,gw20i} considers the fact that the bottomonia are not produced at rest, but with an average finite transverse momentum of $\langle p_T \rangle\simeq 5-6$ GeV.
%{and vanishing azimuthal anisotropy}.
%with corresponding average velocities $\langle v \rangle\simeq 0.46–0.54$ \cite{ngw13}. 
Hence, the relativistic Doppler effect of the moving bottomonia in the hot medium of gluons and light quarks must be considered \cite{hnw17}, a static limit is not applicable.
%In particular, the tiny difference in sequential suppression between the $\Upsilon(2S)$ and the $\Upsilon(3S)$ states appears to require a shorter
%formation time for the least-bound state.

Since the local thermalization time for gluons in a relativistic heavy-ion collision at LHC energies is of the order of 0.1 fm \cite{blmt17,fmr18,gw22,rgw26} and thus, shorter than the formation times for all heavy-quark bound states of interest,
 the expanding and cooling background medium of gluons and light $u$-,\,$d$-,\,$s$-quarks can safely be treated in the hydrodynamic approximation, as has been done in most of the theoretical works that consider the in-medium dissociation of quarkonia such as \cite{brambilla-etal-2011,strickland-2011,song12,peng15,jpb16,rapp17a,fela18,strickland-2019,rot20,
gw20i,bram23}. We shall also use it in the present investigation
%, with an initiation time  of $\tau_\mathrm{ini}=0.1$ fm for the expanding medium, and 
with centrality-dependent initial conditions that depend on the number of both, participants and binary collisions \cite{gw20i}.

Our model for the transverse-momentum and centrality-dependent in-medium suppression of bottomonia states is briefly reviewed in the next section. The determination of the model parameters initial central temperature $T_0$ and $\Upsilon(1S,2S,3S)$ formation times $\tau_\mathrm{F}^{nS}$ in two-dimensional $\chi^2$ minimizations with respect to recent CMS data \cite{cms24} is considered in  Section\,3.
Conclusions are drawn in Section\,4.

%The CGC initial states for the central-rapidity source based on $k_{T}$-factorization, and  for the two fragmentation sources based on hybrid factorization are given in the next section. The diffusion-model approach to the subsequent time evolution of the fragmentation distribution functions in rapidity space and the numerical solution of the corresponding FPE with the CGC initial conditions is considered in Section\,3. Pseudorapidity distributions for produced particles are discussed in Section\,4,  with results for charged-hadron production for $p$-O collisions at $\sqrt{s_{NN}}=9.618$ TeV for seven centrality classes and minimum bias presented in Section\,5. Conclusions are given in Section\,6.
\section{The model}
\label{initial}
Our model accounts for the dissociation of bottomonia states in the hot medium of thermalized gluons and light $u$-,\,$d$-,\,$s$-quarks \cite{gw20i}. It causes a suppression
of these states compared to the expectation from $pp$ collisions at the same energy, scaled with the number of binary collisions. For bottomonium,
six states below the open-bottom threshold are relevant for the dissociation and feed-down processes, namely, $\Upsilon(1S,2S,3S)$ and $\chi_b(1P,2P,3P)$, where we average over hyperfine states. In \cite{ngw14,hnw17,dhw19} we have also included the effect of the running  coupling with energy. This reduces the temperature dependence of the rms radii of the six states under investigation.

We obtain the corresponding color-singlet wavefunctions $\psi_{nlm}(r,\theta,\varphi,T) = g_{nl}(r,T) Y_{lm}(\theta,\varphi)/r$ by solving the radial stationary Schr\"odinger equation for temperatures $T=0-600$ MeV
\begin{equation}
 \partial_r^2 g_{nl}(r,T) =  m_b\, \bigg( V_{\mathrm{eff}}^{nl}(r,T) - E_{nl}(T) + \frac{i\Gamma_{\mathrm{damp}}^{nl}(T)}{2} \bigg)\, g_{nl}(r,T)	\label{radialschroedinger}
\end{equation}
where $\Gamma_{\mathrm{damp}}^{nl}(T)$ is the damping width of the state $|nl\rangle$, $E_{nl}$ the binding energy, $m_b$ the bottom mass and $V_{\mathrm{eff}}^{nl}$ an effective interaction potential that contains the centrifugal barrier and a complex interaction potential $V_{nl}$
%, whose real part vanishes at infinity due to screening,
\begin{equation}
 V_{\mathrm{eff}}^{nl}(r,T) =  V_{nl}(r,T) + \frac{l(l+1)}{m_b r^2}\,,
	\end{equation}
	\begin{equation}
 V_{nl}(r,T) = - \frac{\sigma}{m_D(T)} e^{-m_D(T)r}
	- C_F \alpha_{nl}(T) \Big( \frac{e^{-m_D(T)r}}{r} + iT \phi(m_D(T)r) \Big)
		  \label{complex-potential}
	\end{equation}
	with the temperature-dependent {perturbative hard thermal loop Debye mass {\cite{brapi90,reb93}}} that defines the screening 
	\begin{equation}
 {m_D(T) = T \sqrt{4\pi \alpha_s(T) \frac{2N_c + N_f}{6}}}\,,	
 \label{debyemass}
\end{equation}
and the imaginary part  \cite{laine07} that is proportional to the temperature $T$, and defined through
\begin{equation}
	 \phi(x) = \int\limits_0^\infty \frac{dz \, 2z}{(1+z^2)^2} \left( 1 - \frac{\sin xz}{xz} \right)\,.
	  \label{imaginary-potential}
	 \end{equation}
	 %Lattice results such as \cite{buro17} still deviate from this form of the imaginary part.
The string tension \cite{ja86} equals $\sigma = 0.192$ GeV$^2${.} The complex potential Eq.\,(\ref{complex-potential}) is thus a combination of the potential \cite{laine07,brambilla-etal-2008,beraudo-etal-2008} and the non-perturbative potential ansatz from \cite{karsch-etal-1988}.
The number of colors and flavors, respectively, are set to $N_c = N_f = 3$ and $C_F=(N_c^2-1)/(2N_c)$. The variable $\alpha_{nl}$ denotes the strong coupling $\alpha_s$ evaluated at the soft scale $S_{nl}(T)$,
\begin{equation}
 \alpha_{nl}(T) = \alpha_s(S_{nl}(T))\,, \quad S_{nl}(T) = \langle 1/r \rangle_{nl}(T)\,.	
 \label{coupling}
\end{equation}
We have used the one-loop expression for the running of the coupling. Using values for $\alpha_s$ with matched charm- and bottom-masses \cite{bethke-2013} yields the QCD-scale $\Lambda_\mathrm{QCD} = 276.3$ MeV. Since the coupling constant $\alpha_{nl}(T)$ depends on the solution,
an iterative procedure allows us to solve Eq.\,(\ref{radialschroedinger}) with a \texttt{C++} code \cite{ngw14}. 

The combined action of color screening through the real part of the quark-antiquark potential, and collisional damping through its imaginary part has a substantial effect on the screening and damping of the excited states, whereas the $\Upsilon(1S)$ spin-triplet ground state is less affected (except for the most central collisions) due to its large binding energy of 1.1 GeV with respect to the open-bottom threshold.

 For sufficiently large gluon energies $E_g>|E_{nl}|$, however, the bottomonia states can be dissociated directly in a singlet to octet transition. This gluodissociation is treated separately from damping due to the separation of scales \cite{bram08,brambilla-etal-2011}. We have calculated the dissociation cross section \cite{bgw12,ngw13}, and folded it with a Bose--Einstein distribution for the gluons to obtain the temperature-dependent dissociation widths $\Gamma_\mathrm{diss}^{nl}(T)$ for the six states of interest.
For each spacetime point $(t,x^1,x^2)$ in the transverse plane and for every temperature $T(t)$ we calculate the total dissociation widths of the six screened bottomonia states from
 the incoherent sum of damping and dissociation widths,
\begin{equation}
\Gamma_\mathrm{tot}^{nl}(t,x^1,x^2)=\Gamma_\mathrm{damp}^{nl}(t,x^1,x^2)+\Gamma_\mathrm{diss}^{nl}(t,x^1,x^2)\,.
\end{equation}
 Dissociation through screening of the real part of the quark-antiquark potential is considered by setting the total decay width to infinity once the energy eigenvalue of a state $|nl\rangle$ meets the continuum threshold.
 
Both damping and gluodissociation occur during collective expansion and cooling of the hot medium. The expansion velocity of the medium generally differs from the rms velocity of the bottomonia, which have been created with a finite transverse momentum $\langle p_T\rangle\simeq 5-6$ GeV. Due to the large bottom mass, it does not change much when bottomonia pass through the medium. Therefore, we take the relativistic Doppler effect \cite{esco13} into account \cite{hnw17,gw20i} when computing the temperature-dependent dissociation widths.

Our hydrodynamic treatment of the background bulk evolution with longitudinal expansion has been described in detail earlier \cite{ngw13,hnw17}. We model the QGP as a relativistic, perfect fluid consisting of gluons and massless up-, down- and strange-quarks and solve the corresponding equations in the longitudinally co-moving frame. {This earlier model now includes an additional transverse expansion} \cite{ngw14,gw20i}, which causes additional cooling, resulting in less pronounced dissociation. In line with the gluon thermalization timescale, we use an initiation time  of $\tau_\mathrm{ini}=0.1$ fm for the expanding medium. Including viscosity would slow down the expansion, cause a slight increase in the damping and dissociation widths, and a corresponding decrease in the initial central temperature $T_0$ when comparing to data -- without changing the overall picture. When including viscosity, a larger initiation time is usually chosen (typically 0.3-0.6 fm \cite{ch24}) to avoid excessive suppression, thus compensating for the slower time evolution.

The amount of in-medium suppression of bottomonia with transverse momentum $p_T$ for A-A collisions in the centrality bin $c$, where $b_c \leq b < b_{c+1}$, is quantified by the QGP-suppression factors $R^\mathrm{QGP}_{AA,nl}(c,p_T)$. This factor is not directly measurable, because it accounts only for the amount of suppression inside the fireball due to the three processes color screening, collisional damping and gluodissociation. It is given by the ratio of the number of bottomonia that have survived the fireball to the number of bottomonia produced in the collision. The latter scales with the number of binary collisions at a given point in the transverse plane and hence, with the nuclear overlap $T_{AA}$, %$N_{b\bar{b}} $
$N_{\Upsilon,\,\chi_b}(nl)\propto N_\mathrm{coll} \propto T_{AA}$. 

Thus, we define the bottomonium suppression factor in the QGP at centrality $c$ and transverse momentum $p_T$
% $R_{AA,nl}^\mathrm{QGP}$ as 
\begin{equation}
\label{RAAQGP}
 R_{AA,nl}^\mathrm{QGP}(c,p_T)
 =\frac{\int_{b_c}^{b_{c+1}} \mathrm{d}b \, b \int \mathrm{d}^2x \, T_{AA}(b,x^1,x^2) \, D_{nl}(b,p_T,x^1,x^2)}{\int_{b_c}^{b_{c+1}} \mathrm{d}b \, b \int \mathrm{d}^2x \, T_{AA}(b,x^1,x^2)}\,.
\end{equation}
The damping factor $D_{nl}$ is given by the temporal integral over the total bottomonium decay width $\Gamma_{nl}$,
%$b\bar{b}$ decay width $\Gamma_{nl}$,
\begin{equation}
\label{damping-factor}
 D_{nl}(b,p_T,x^1,x^2)
 = \exp\left[ - \int\limits_{\tau_{\mathrm{F}}^{nl} \gamma_{{T},\,nl}(p_T)}^\infty \frac{\mathrm{d}\tau \, \Gamma_{nl}(b,p_T,\tau,x^1,x^2)}{\gamma_{{T},nl}(p_T)}\right],
\end{equation}
%\note[GW]{Variable in Gamma einfuegen?}
where $\tau_{\mathrm{F}}^{nl}$ is the formation time in the bottomonium rest-frame, $\gamma_{{T},\,nl}(p_T) = \sqrt{1 + (p_T/M^{\mathrm{vac}}_{nl})^2}$ the Lorentz-factor due to transverse motion in the longitudinally comoving frame, and $M^{\mathrm{vac}}_{nl}$ the experimentally measured bottomonium vacuum mass. Here, the weighted damping factor $T_{AA} D_{nl}$  scales with the number of surviving bottomonia in the transverse plane $(x^1,x^2)$.

Once the bottomonia states have survived the dissociation processes in the hot quark-gluon plasma, we consider the feed-down cascade from the excited states to the ground state  \cite{vaccaro-etal-2013}. The corresponding branching ratios are taken from the particle data group \cite{pdg24}, and from theory \cite{daghighian-silverman-1987}
for the $\chi_b(3P)$ state.
Due to the rapid melting or depopulation of the excited states caused by the mechanisms in the QGP-phase, the feed-down to the ground state is reduced, resulting in additional 
$\Upsilon(1S)$-suppression with respect to the situation in $pp$~collisions at the same energy. For the excited states, in contrast, the reduction of the feed-down does not substantially modify the suppression factors. The calculated suppression factors for $\Upsilon(1S,2S,3S)$ that include feed-down can then be compared to data.

To estimate the initial populations $N^\text{i}_{AA,nl}$ of the six bottomonia states that we treat explicitly, we consider the measured final populations $N^\text{f}_{{pp},nl}$ of the three $\Upsilon(nS)$-states in $pp$ collisions at the same energy and calculate the decay cascade~\cite{vaccaro-etal-2013} backwards to obtain the initial populations in $pp$, $N^\text{i}_{{pp},nl}$. 
These are then scaled by the number of binary collisions $N_\text{coll}$ yielding the initial populations $N^\text{i}_{AA,nl}$ in the heavy-ion case. 
%$N^\text{i}_{AA,nl} = N_\text{coll} N^\text{i}_{\text{pp},nl}$. 
When the suppression factors are calculated, the number of binary collisions cancels out by definition.
The required branching ratios are taken from the Review of Particle Physics~\cite{pdg24} or from theory where no experimental values are available, as is the case of 
$\chi_b(3P)$. 

Given the initial $\Upsilon(nS)$ populations from $pp$ collisions, the model as outlined above has the initial central temperature $T_0$, and the formation times 
%$\tau_\mathrm{F}^{nS,nP}$ 
$\tau_\mathrm{F}^{nl}$ of the bottomonia states as free parameters. In our previous comparisons \cite{ngw14,hnw17} with centrality- and transverse-momentum-dependent Pb-Pb data, we had integrated the initial populations
over transverse momentum and rapidity, assumed the same formation time $\tau_\mathrm{F}=0.4$ fm for all six states, and adapted the initial central temperature to the data.
For $\sqrt{s_{NN}}=2.76$ TeV Pb-Pb, this resulted in $T_0=480$ MeV \cite{hnw17} when compared to CMS data \cite{cms17}. At higher energies such as $\sqrt{s_{NN}}=5.02$ TeV,
we had extrapolated  $T_0$ using the scaling relation between the initial entropy density and the charged particle multiplicity per unit rapidity, $s_0 \propto dN_\text{ch}/d\eta$ \cite{bjorken-1983,baym-etal-1983,gyulassi-matsui-1984}, and inserted $s_0 \propto T_0^3$ together with measured results for $dN_\text{ch}/d\eta$  \cite{alice17} to obtain an increase by 6.8\% in the initial temperature, $T_0=513$ MeV. This prediction \cite{hnw17} resulted in agreement with CMS data \cite{cms19,cms24} for $\Upsilon(1S)$, but still insufficient suppression for $\Upsilon(2S,3S)$. Our related prediction for $\Upsilon(1S,2S)$ in Au-Au at the much lower RHIC energy of $\sqrt{s_{NN}}=200$ MeV with $T_0=356$ MeV also resulted in insufficient suppression \cite{star23}, likely because the $T_0$ estimate from the entropy argument over such a large energy range is unreliable -- a calculation with $T_0=420$ MeV yields satisfactory agreement with the STAR data. 
 \begin{figure}[ht] 
\centering 
\includegraphics[width=\columnwidth]{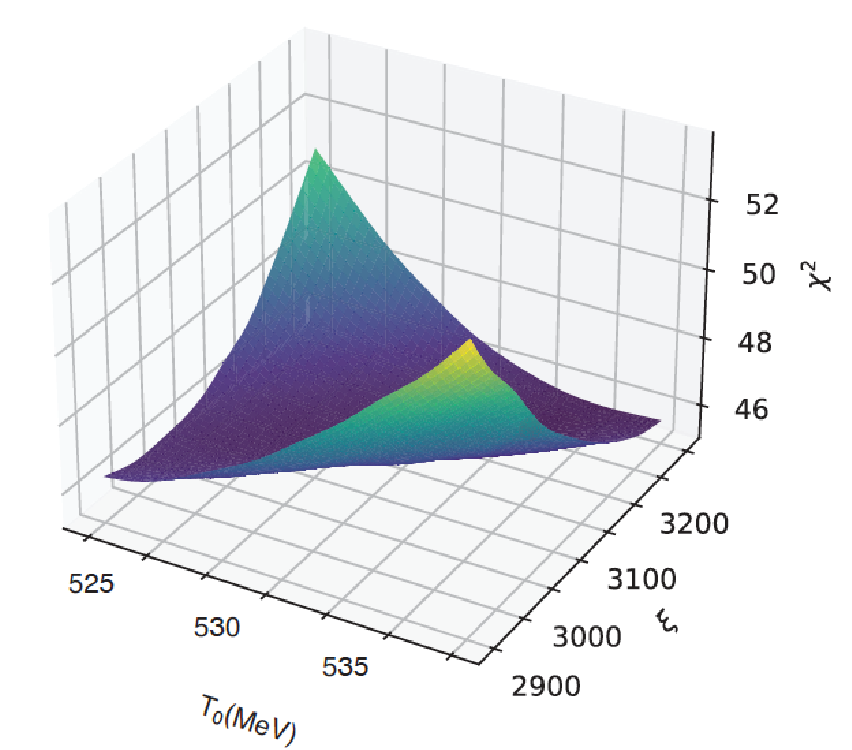} 
\caption{Simultaneous $\chi^2$ optimization of initial central temperature $T_0$ and $\Upsilon(nS)$ formation times in $\sqrt{s_{NN}}=5.02$ TeV Pb-Pb collisions with respect to recent $p_T$- and centrality-dependent CMS data \cite{cms24}. The minimum is found at $T_0=535.9$ MeV, the minimal $\xi$ value corresponds to an average $\Upsilon(1S)$ formation time $\langle\tau_\mathrm{F}^{1S}\rangle\simeq {0.34}$ fm, with $\chi^2={45.6}$ and {$\chi^2/$ndf=45.6/32=1.43.}}
\label{fig1}
\end{figure}
 \begin{figure}[ht] 
\centering 
\includegraphics[width=0.9\columnwidth]{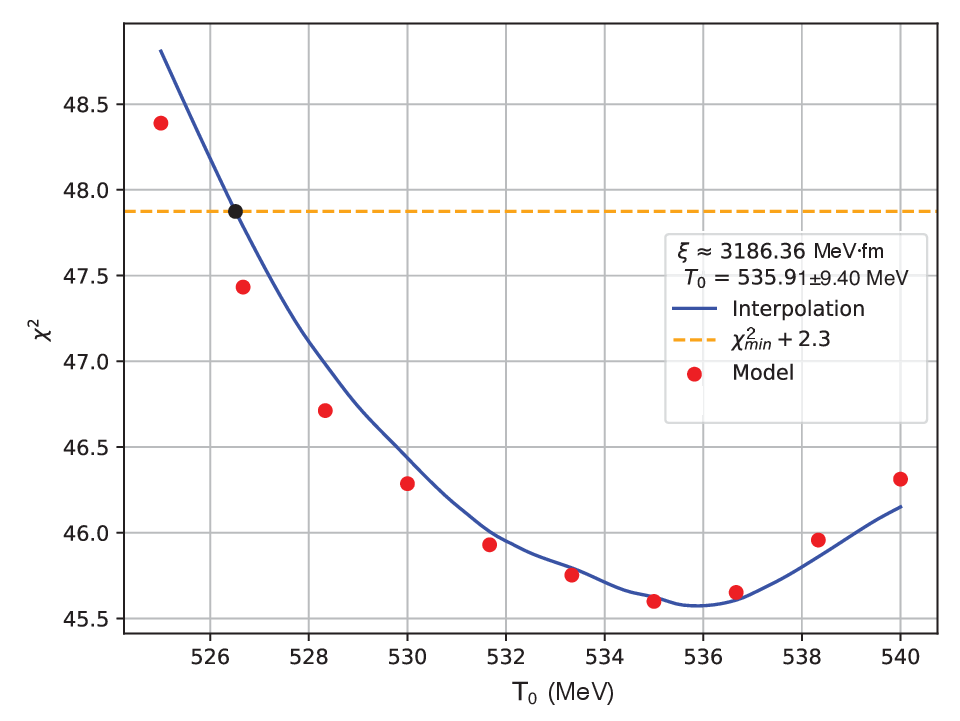} 
\caption{Cut of the 3D optimisation plot shown in Fig.\,1 along the temperature axis through the minimum. The dashed line corresponds to the $68\%$ confidence interval ($1\sigma$) \cite{pre93}.} 
\label{fig2}
\end{figure}
\section{Model calculations for Pb-Pb at LHC energies}
\label{pbpb}
The newly available precise CMS data for centrality- and $p_T$-dependent suppression of $\Upsilon(1S,2S,3S)$ in $\sqrt{s_{NN}}=5.02$ TeV Pb-Pb collisions \cite{cms24} offer the possibility to determine our model parameters $T_0$ and $\tau_\mathrm{F}$ in $\chi^2$ minimizations with respect to these data. Different from our earlier predictions, we now allow for state-dependent bottomonium formation times, $\tau_\mathrm{F}\rightarrow\tau_\mathrm{F}^{nl}$.

Regarding the initial populations of the bottomonia states that we obtain through an inverse cascade calculation from measured distributions in $pp$ collisions at the same energy, we start from the transverse-momentum dependent $pp$ data that are now available at 5.02 TeV from CMS \cite{cms19}, LHCb \cite{lhcb23} and ATLAS \cite{atlas23}, albeit in differing transverse-momentum intervals {and rapidity bins}. The measured occupation of the $\Upsilon(nS)$ states rises with increasing binding energy -- except for the $\Upsilon(3S)$ data from CMS, which show a slightly larger population than $\Upsilon(2S)$ in the highest transverse-momentum bin, $p_T\le30$ GeV. 

When using these CMS $pp$ data in our model calculation for Pb-Pb at 5.02 TeV, the $\Upsilon(3S)$ state therefore turns out to be less suppressed than $\Upsilon(2S)$ { in the highest transverse-momentum bin}, contrary to the sequential pattern seen in the CMS Pb-Pb data { for all trtansverse momenta}. Hence, we turn to the LHCb $pp$ data \cite{lhcb23} with higher momentum resolution to obtain the initial populations of the bottomonia states. As a disadvantage, these data are only available up to $p_T\le20$ GeV, and the rapidity interval differs from CMS -- the rapidity dependence is, however, very weak: { CMS found it to be negligeable within the experimental error bars for  $|y| <2.4$ \cite{cms19}}.
% and we do not expect a substantially different behaviour for the LHCb rapidity range $2.0 <|y| <4.5$.}

 \begin{figure}[ht] 
\centering 
\includegraphics[width=\columnwidth]{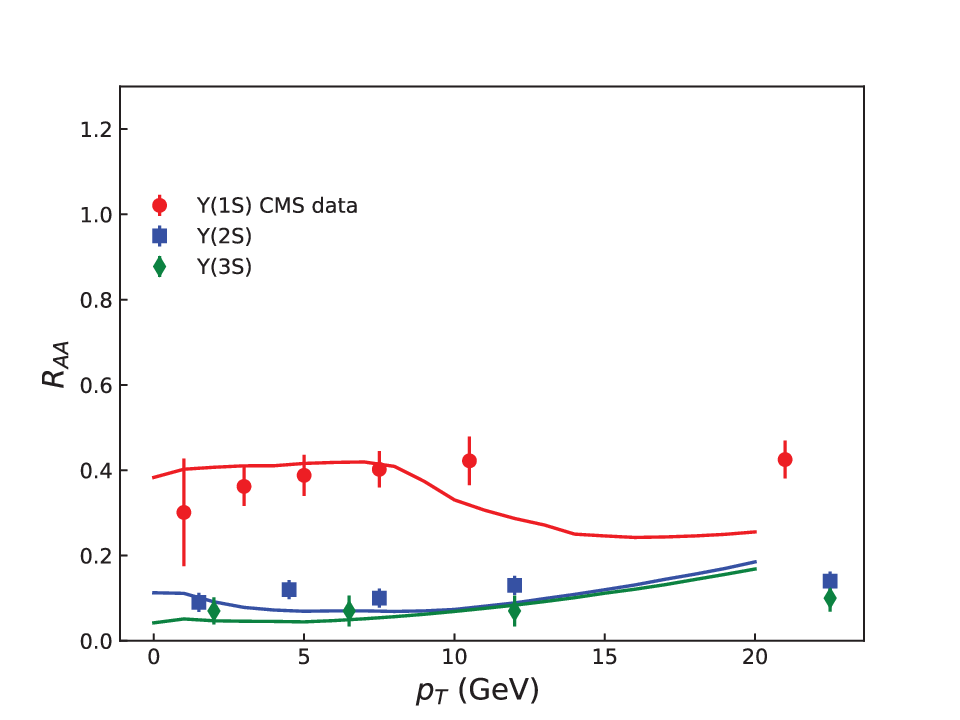} 
\caption{Calculated nuclear suppression factors $R_{AA}$  of the $\Upsilon(1S,2S,3S)$ states as functions of transverse momentum $p_T$ for $|y|<2.4$ in $0-90\%$ $\sqrt{s_{NN}}=5.02$ TeV Pb-Pb collisions (solid curves) compared with recent CMS data (symbols) \cite{cms24}. The $\Upsilon(1S)$ data are from \cite{cms19}.  
The model parameters initial central temperature $T_0=535.9$ MeV and respective formation times $\tau_\mathrm{F}= {0.34, 0.32, 0.31}$ fm of the three $\Upsilon(nS)$ states  are from our simultaneous $\chi^2$ optimisation shown in Figs.\,1,2. } 
\label{fig3}
\end{figure}

With respect to the formation times of the six bottomonia states involved, 
 different approaches to determine them in vacuum and in the medium have been considered in the past. For example, $\tau_\mathrm{F}$ has been defined as the time when the separation of the quark-antiquark pair attains the size of a physical quarkonium state. Whereas this may be true in vacuum, in the hot medium the temperature-dependent spatial extent of the quarkonium states is modified due to the running coupling \cite{ngw14}, and a proportionality of the formation time to the inverse binding energy may become invalid. Indeed, the 
excited $\Upsilon(2S,3S)$ states may have shorter in-medium formation times than the spin-triplet $\Upsilon(1S)$ ground state -- a situation that has a low-energy correspondence in the formation of muonic atoms, where the excited states also have smaller formation times than the ground state inspite of the larger spatial extent of their wavefunctions \cite{lwh74}.
 %\begin{figure}[hb] 
 \begin{figure}
\centering 
\includegraphics[width=\columnwidth]{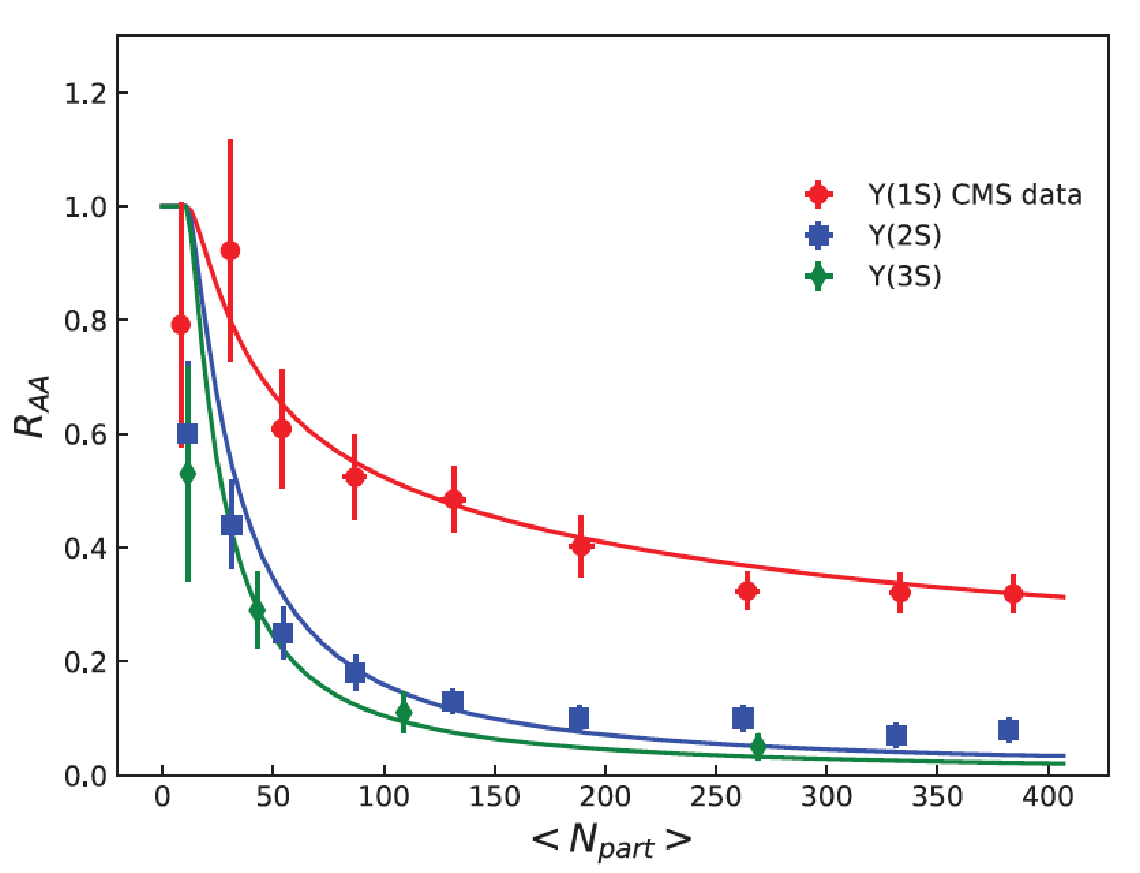} 
\caption{Calculated nuclear suppression factors $R_{AA}$ of the $\Upsilon(1S,2S,3S)$ states as functions of centrality (average number of participants $\langle N_\mathrm{part}\rangle$) in $\sqrt{s_{NN}}=5.02$ TeV Pb-Pb collisions (solid curves) for $|y|<2.4$ and {$p_T<20$} GeV compared with recent CMS data (symbols) \cite{cms24}. The $\Upsilon(1S)$ data are from \cite{cms19}.    
The model parameters initial central temperature $T_0=535.9$ MeV and respective formation times $\langle \tau_\mathrm{F}^{nS}\rangle= {0.34, 0.32, 0.31}$ fm of the three $\Upsilon(nS)$ states  are from our simultaneous $\chi^2$ optimisation shown in Figs.\,1,2. } 
\label{fig4}
\end{figure}
We have checked the proposition of formation times being inversely proportional to the binding energies of the six bottomonia states, $\tau_\mathrm{F}^{nl}\propto 1/E_\mathrm{B}^{nl}$
within our model. In a simultaneous $\chi^2$ minimization with respect to the centrality- and $p_T$-dependent CMS Pb-Pb data \cite{cms24}, it is not possible to obtain the correct sequential suppression of the $\Upsilon(2S)$ and $\Upsilon(3S)$ states, because the lifetime of the weakly bound states in the hot medium is too short (their formation times being too long) to become sufficiently suppressed. 

Instead, we propose a two-dimensional parameter optimization with initial central temperature $T_0$ and formation times proportional to the inverse masses of the six states, $\tau_\mathrm{F}^{nl}=\xi/M_{nl}$.
The optimization for 5.02 TeV Pb-Pb yields an initial central temperature $T_0=536$ MeV and formation times defined by $\xi$ as shown in Fig.\,\ref{fig1}, with a cut along the temperature axis in Fig.\,\ref{fig2}. The corresponding average $\Upsilon(1S,2S,3S)$ formation times  $\langle\tau_\mathrm{F}^{nS}\rangle=$ {0.34 fm, 0.32 fm, 0.31 fm} are consistent with an early model-independent determination of the $\Upsilon(1S)$ vacuum formation time from $e^+e^-$ annihilation that yielded 0.32 fm \cite{kt99}.
% -- the in-medium formation times must certainly be larger than the vacuum value.
{The optimized formation-time values remain close to $\langle\tau_\mathrm{F}\rangle=0.3$ fm that we had used for all six bottomonia states in our previous predictions \cite{hnw17,gw20i} for $\Upsilon$ suppression in Pb-Pb at LHC energies}.

In Figs.\,\ref{fig3},\ref{fig4}, our results are shown together with the $p_T$- and centrality-dependent CMS data \cite{cms24} for 5.02 TeV Pb-Pb. {The $p_T$-dependent data for all three observed bottomonia states agree with our model within the error bars up to $p_T\le10$ GeV, Fig.\,\ref{fig3}. The calculated $R_{AA}$ increases beyond $p_T\simeq 15$ GeV for all three $\Upsilon$ states, since high-momentum bottomonia spend less time in the medium and therefore, suffer less suppression.}
Because we obtained the initial $\Upsilon$ populations from LHCb $pp$ data \cite{lhcb23} that are only available up to transverse momenta $p_T\le 20$ GeV, we show Pb-Pb results up to this value of transverse momentum. {The data tend to remain quite flat at high transverse momenta, but the bins are relatively wide, and we will have to rely on future more precise results at high transverse momenta}.

As is the case for the momentum-dependent results, the centrality-dependent suppression factors shown in Fig.\,\ref{fig4} have the correct sequential suppression pattern and agree with the data. The simultaneous two-dimensional fit of both data sets has $\chi^2/\mathrm{ndf}=45.6/32=1.43$.

%\cite{gw20i,fmr18,star23,rgw26,bgw12,vaccaro-etal-2013,ngw13,ngw14,hnw17,dhw19,pre93}

\section{Conclusions}
\label{concl}
We have used recent centrality- and transverse-momentum-dependent bottomonia suppression data for  $\sqrt{s_{NN}}=5.02$ TeV Pb-Pb collisions from the CMS collaboration to
determine the parameters initial central temperature $T_0$ and bottomonia formation times $\tau_\mathrm{F}^{nl}$ within our model for quarkonia suppression from the data in a simultaneous two-dimensional $\chi^2$ minimization. The resulting value  $T_0=536$ MeV is higher than our estimate from entropy arguments ($T_0=513$ MeV) in an earlier prediction for the $\Upsilon(nS)$ suppression in 5.02 TeV Pb-Pb, where we had used initial populations from $pp$ data that were integrated over rapidity and transverse momentum. 

The in-medium formation times {are found to be}  $\langle\tau_\mathrm{F}^{nS}\rangle=$ {0.34 fm, 0.32 fm, 0.31 fm}  for the observed $\Upsilon(1S,2S,3S)$ states. {For the $\Upsilon(1S)$ state}, this is larger than the model-independent vacuum formation time $\tau=0.32$ fm, and much larger than
% the initiation time for the hydrodynamic expansion, which we take as $\tau_\mathrm{ini}=0.1$ fm, in agreement with 
the gluon thermalization timescale of 0.1 fm. Since the total interaction time for central Pb-Pb collisions at LHC energies is of the order of $8-10$ fm, the deduced formation-time values are sufficiently short to provide the observed suppression of the $\Upsilon(nS)$ states, and also the correct sequential suppression pattern.  

Once more precise data for $\Upsilon(nS)$ suppression in $\sqrt{s_{NN}}=200$ GeV  Au-Au collisions become available, a corresponding $\chi^2$ minimization of initial central temperature and formation times can also be performed at RHIC energies.
\section*{Acknowledgement}
We are grateful to Johannes H{\"o}lck (now at Landesbank Stuttgart, Germany) for discussions, and for his cooperation with JM in handling the \texttt{C++} code that is the basis for our determination of initial central QGP temperature, and in-medium $\Upsilon$ formation times from CMS data.
%\newpage
%\begin{thebibliography}{10}
%\newpage
\bibliographystyle{elsarticle-num-names}
%\bibitem{}
%\bibliography{gw_26}

%\bibliography{gw_26}

\begin{thebibliography}{56}
\expandafter\ifx\csname natexlab\endcsname\relax\def\natexlab#1{#1}\fi
\providecommand{\url}[1]{\texttt{#1}}
\providecommand{\href}[2]{#2}
\providecommand{\path}[1]{#1}
\providecommand{\DOIprefix}{doi:}
\providecommand{\ArXivprefix}{arXiv:}
\providecommand{\URLprefix}{URL: }
\providecommand{\Pubmedprefix}{pmid:}
\providecommand{\doi}[1]{\href{http://dx.doi.org/#1}{\path{#1}}}
\providecommand{\Pubmed}[1]{\href{pmid:#1}{\path{#1}}}
\providecommand{\bibinfo}[2]{#2}
\ifx\xfnm\relax \def\xfnm[#1]{\unskip,\space#1}\fi
%Type = Article
\bibitem[{Adare et~al.(2007)}]{phe07}
\bibinfo{author}{A.~Adare}, et~al. (\bibinfo{collaboration}{PHENIX
  Collaboration}),
\newblock \bibinfo{title}{${J/\psi}$ production versus centrality, transverse
  momentum, and rapidity in {Au-Au} collisions at $\sqrt{s_{NN}} = 200 \; \rm
  {GeV}$},
\newblock \bibinfo{journal}{Phys. Rev. Lett.} \bibinfo{volume}{98}
  (\bibinfo{year}{2007}) \bibinfo{pages}{0232301}.
%Type = Article
\bibitem[{Abelev et~al.(2012)}]{alice12i}
\bibinfo{author}{B.~Abelev}, et~al. (\bibinfo{collaboration}{ALICE
  Collaboration}),
\newblock \bibinfo{title}{${J/\psi}$ suppression at forward rapidity in {Pb-Pb}
  collisions at $\sqrt{s_{NN}} = 2.76 \; \rm {TeV}$},
\newblock \bibinfo{journal}{Phys. Rev. Lett.} \bibinfo{volume}{109}
  (\bibinfo{year}{2012}) \bibinfo{pages}{072301}.
%Type = Article
\bibitem[{{A}charya et~al.(2020)}]{alice20jhep}
\bibinfo{author}{S.~{A}charya}, et~al. (\bibinfo{collaboration}{ALICE
  Collaboration}),
\newblock \bibinfo{title}{Studies of ${J/\psi}$ production at forward rapidity
  in {Pb-Pb} collisions at $\sqrt{s_{NN}}=5.02$ {TeV}},
\newblock \bibinfo{journal}{JHEP} \bibinfo{volume}{2020} (\bibinfo{year}{2020})
  \bibinfo{pages}{41}.
%Type = Article
\bibitem[{{Matsui} and {Satz}(1986)}]{ms86}
\bibinfo{author}{T.~{Matsui}}, \bibinfo{author}{H.~{Satz}},
\newblock \bibinfo{title}{{$J/\psi$ Suppression by Quark-Gluon Plasma
  Formation}},
\newblock \bibinfo{journal}{Phys. Lett. B} \bibinfo{volume}{178}
  (\bibinfo{year}{1986}) \bibinfo{pages}{416}.
%Type = Article
\bibitem[{Chatrchyan et~al.(2012)}]{CMS-2012}
\bibinfo{author}{S.~Chatrchyan}, et~al. (\bibinfo{collaboration}{{CMS
  Collaboration}}),
\newblock \bibinfo{title}{Observation of sequential {Upsilon} suppression in
  {PbPb} collisions},
\newblock \bibinfo{journal}{Phys. Rev. Lett.} \bibinfo{volume}{109}
  (\bibinfo{year}{2012}) \bibinfo{pages}{{222301}}.
%Type = Article
\bibitem[{Khachatryan et~al.(2017)}]{cms17}
\bibinfo{author}{V.~Khachatryan}, et~al. (\bibinfo{collaboration}{CMS
  Collaboration}),
\newblock \bibinfo{title}{Suppression of {$\Upsilon(1S)$}, {$\Upsilon(2S)$},
  and {$\Upsilon(3S)$} quarkonium states in {PbPb} collisions at $\sqrt{s} =
  2.76$ {TeV}},
\newblock \bibinfo{journal}{Phys. Lett. B} \bibinfo{volume}{770}
  (\bibinfo{year}{2017}) \bibinfo{pages}{357}.
%Type = Article
\bibitem[{Aboona et~al.(2023)}]{star23}
\bibinfo{author}{B.~E. Aboona}, et~al. (\bibinfo{collaboration}{STAR
  Collaboration}),
\newblock \bibinfo{title}{Measurement of sequential ${\Upsilon}$ suppression in
  {Au+Au} collisions at $\sqrt{s_{NN}} = 200 \; \rm {GeV}$ with the {STAR}
  experiment},
\newblock \bibinfo{journal}{Phys. Rev. Lett.} \bibinfo{volume}{130}
  (\bibinfo{year}{2023}) \bibinfo{pages}{112301}.
%Type = Article
\bibitem[{Adam et~al.(2017)}]{alice17i}
\bibinfo{author}{J.~Adam}, et~al. (\bibinfo{collaboration}{ALICE
  Collaboration}),
\newblock \bibinfo{title}{${J/\psi}$ suppression at forward rapidity in {Pb-Pb}
  collisions at $\sqrt{s_{NN}} = 5.02 \; \rm {TeV}$},
\newblock \bibinfo{journal}{Phys. Lett. B} \bibinfo{volume}{766}
  (\bibinfo{year}{2017}) \bibinfo{pages}{212--224}.
%Type = Article
\bibitem[{Acharya et~al.(2024)}]{alice24}
\bibinfo{author}{S.~Acharya}, et~al. (\bibinfo{collaboration}{ALICE
  Collaboration}),
\newblock \bibinfo{title}{Prompt and non-prompt ${J/\psi}$ production at
  midrapidity in {Pb-Pb} collisions at $\sqrt{s_{NN}} = 5.02 \; \rm {GeV}$},
\newblock \bibinfo{journal}{JHEP} \bibinfo{volume}{02} (\bibinfo{year}{2024})
  \bibinfo{pages}{066}.
%Type = Article
\bibitem[{Sirunyan et~al.(2019)}]{cms19}
\bibinfo{author}{A.~M. Sirunyan}, et~al. (\bibinfo{collaboration}{CMS
  Collaboration}),
\newblock \bibinfo{title}{Measurement of nuclear modification factors of
  {$\Upsilon(1S)$, $\Upsilon(2S)$, and $\Upsilon(3S)$} mesons in {PbPb}
  collisions at $\sqrt{s_{NN}} = 5.02$ {TeV}},
\newblock \bibinfo{journal}{Phys. Lett. B} \bibinfo{volume}{790}
  (\bibinfo{year}{2019}) \bibinfo{pages}{270--293}.
%Type = Article
\bibitem[{Tumasyan et~al.(2024)}]{cms24}
\bibinfo{author}{A.~Tumasyan}, et~al. (\bibinfo{collaboration}{CMS
  Collaboration}),
\newblock \bibinfo{title}{Observation of the {$\Upsilon(3S)$} meson and
  suppression of {{$\Upsilon$}} states in {Pb-Pb} collisions at $\sqrt{s_{NN}}
  = 5.02 \; \rm {TeV}$},
\newblock \bibinfo{journal}{Phys. Rev. Lett.} \bibinfo{volume}{133}
  (\bibinfo{year}{2024}) \bibinfo{pages}{022302}.
%Type = Article
\bibitem[{Hoelck et~al.(2017)Hoelck, Nendzig, and Wolschin}]{hnw17}
\bibinfo{author}{J.~Hoelck}, \bibinfo{author}{F.~Nendzig},
  \bibinfo{author}{G.~Wolschin},
\newblock \bibinfo{title}{In-medium {{$\Upsilon$}} suppression and feed-down in
  {UU and PbPb} collisions},
\newblock \bibinfo{journal}{Phys. Rev. C} \bibinfo{volume}{95}
  (\bibinfo{year}{2017}) \bibinfo{pages}{024905}.
%Type = Article
\bibitem[{Wolschin(2020)}]{gw20i}
\bibinfo{author}{G.~Wolschin},
\newblock \bibinfo{title}{Bottomonium spectroscopy in the quark-gluon plasma},
\newblock \bibinfo{journal}{Int. J. Mod. Phys. A} \bibinfo{volume}{35}
  (\bibinfo{year}{2020}) \bibinfo{pages}{2030016}.
%Type = Article
\bibitem[{Song et~al.(2015)Song, Ko, and Lee}]{ko15}
\bibinfo{author}{T.~Song}, \bibinfo{author}{C.~M. Ko}, \bibinfo{author}{S.~H.
  Lee},
\newblock \bibinfo{title}{{Quarkonium formation time in relativistic heavy-ion
  collisions}},
\newblock \bibinfo{journal}{Phys. Rev. C} \bibinfo{volume}{91}
  (\bibinfo{year}{2015}) \bibinfo{pages}{044909}.
%Type = Article
\bibitem[{Chen et~al.(2024)Chen, Chen, and Zhao}]{ch24}
\bibinfo{author}{G.~Chen}, \bibinfo{author}{B.~Chen},
  \bibinfo{author}{J.~Zhao},
\newblock \bibinfo{title}{Bottomonium evolution with in-medium heavy quark
  potential from lattice {QCD}},
\newblock \bibinfo{journal}{Eur. Phys. J. C} \bibinfo{volume}{84}
  (\bibinfo{year}{2024}) \bibinfo{pages}{869}.
%Type = Article
\bibitem[{Acharya et~al.(2019)}]{alice19b}
\bibinfo{author}{S.~Acharya}, et~al. (\bibinfo{collaboration}{ALICE
  Collaboration}),
\newblock \bibinfo{title}{${\Upsilon}$ suppression at forward rapidity in
  {Pb-Pb} collisions at $\sqrt{s_{NN}}=5.02$ {TeV}},
\newblock \bibinfo{journal}{Phys. Lett. B} \bibinfo{volume}{790}
  (\bibinfo{year}{2019}) \bibinfo{pages}{89}.
%Type = Article
\bibitem[{{Nendzig} and {Wolschin}(2013)}]{ngw13}
\bibinfo{author}{F.~{Nendzig}}, \bibinfo{author}{G.~{Wolschin}},
\newblock \bibinfo{title}{{{$\Upsilon$}} suppression in {PbPb} collisions at
  energies available at the {CERN Large Hadron Collider}},
\newblock \bibinfo{journal}{Phys. Rev. C} \bibinfo{volume}{87}
  (\bibinfo{year}{2013}) \bibinfo{pages}{024911}.
%Type = Article
\bibitem[{Blaizot et~al.(2017)Blaizot, Liao, and Mehtar-Tani}]{blmt17}
\bibinfo{author}{J.-P. Blaizot}, \bibinfo{author}{J.~Liao},
  \bibinfo{author}{Y.~Mehtar-Tani},
\newblock \bibinfo{title}{The thermalization of soft modes in non-expanding
  isotropic quark gluon plasmas},
\newblock \bibinfo{journal}{Nucl. Phys. A} \bibinfo{volume}{961}
  (\bibinfo{year}{2017}) \bibinfo{pages}{37--67}.
%Type = Article
\bibitem[{Florkowski et~al.(2018)Florkowski, Maksymiuk, and Ryblewski}]{fmr18}
\bibinfo{author}{W.~Florkowski}, \bibinfo{author}{E.~Maksymiuk},
  \bibinfo{author}{R.~Ryblewski},
\newblock \bibinfo{title}{Coupled kinetic equations for fermions and bosons in
  the relaxation-time approximation},
\newblock \bibinfo{journal}{Phys. Rev. C} \bibinfo{volume}{97}
  (\bibinfo{year}{2018}) \bibinfo{pages}{024915}.
%Type = Article
\bibitem[{Wolschin(2026)}]{gw22}
\bibinfo{author}{G.~Wolschin},
\newblock \bibinfo{title}{Nonlinear diffusion of gluons},
\newblock \bibinfo{journal}{Physica A} \bibinfo{volume}{597}
  (\bibinfo{year}{2026}) \bibinfo{pages}{127299}.
%Type = Article
\bibitem[{R{\"o}ssler and Wolschin(2026)}]{rgw26}
\bibinfo{author}{J.~R{\"o}ssler}, \bibinfo{author}{G.~Wolschin},
\newblock \bibinfo{title}{Numerical solution of the nonlinear boson diffusion
  equation for gluons},
\newblock \bibinfo{journal}{Physica A} \bibinfo{volume}{682}
  (\bibinfo{year}{2026}) \bibinfo{pages}{131157}.
%Type = Article
\bibitem[{Brambilla et~al.(2011)Brambilla, Escobedo, Ghiglieri, and
  Vairo}]{brambilla-etal-2011}
\bibinfo{author}{N.~Brambilla}, \bibinfo{author}{M.~A. Escobedo},
  \bibinfo{author}{J.~Ghiglieri}, \bibinfo{author}{A.~Vairo},
\newblock \bibinfo{title}{{Thermal width and gluo-dissociation of quarkonium in
  pNRQCD}},
\newblock \bibinfo{journal}{JHEP} \bibinfo{volume}{2011} (\bibinfo{year}{2011})
  \bibinfo{pages}{116}.
%Type = Article
\bibitem[{Strickland(2011)}]{strickland-2011}
\bibinfo{author}{M.~Strickland},
\newblock \bibinfo{title}{{Thermal {$\Upsilon$}(1$s$) and $\chi_{b1}$
  suppression in $\sqrt{s_{NN}} = 2.76 \;\rm TeV$ Pb-Pb collisions at the
  LHC}},
\newblock \bibinfo{journal}{Phys. Rev. Lett.} \bibinfo{volume}{107}
  (\bibinfo{year}{2011}) \bibinfo{pages}{132301}.
%Type = Article
\bibitem[{Song et~al.(2012)Song, Han, and Ko}]{song12}
\bibinfo{author}{T.~Song}, \bibinfo{author}{K.~C. Han}, \bibinfo{author}{C.~M.
  Ko},
\newblock \bibinfo{title}{{Bottomonia suppression in heavy-ion collisions}},
\newblock \bibinfo{journal}{Phys. Rev. C} \bibinfo{volume}{85}
  (\bibinfo{year}{2012}) \bibinfo{pages}{014902}.
%Type = Article
\bibitem[{Liu et~al.(2015)Liu, Zhou, and Zhuang}]{peng15}
\bibinfo{author}{Y.~Liu}, \bibinfo{author}{K.~Zhou},
  \bibinfo{author}{P.~Zhuang},
\newblock \bibinfo{title}{{Quarkonia in high energy nuclear collisions}},
\newblock \bibinfo{journal}{Int. J. Mod. Phys. E} \bibinfo{volume}{24}
  (\bibinfo{year}{2015}) \bibinfo{pages}{1530015}.
%Type = Article
\bibitem[{Blaizot et~al.(2016)Blaizot, Boni, Faccioli, and Garberoglio}]{jpb16}
\bibinfo{author}{J.-P. Blaizot}, \bibinfo{author}{D.~D. Boni},
  \bibinfo{author}{P.~Faccioli}, \bibinfo{author}{G.~Garberoglio},
\newblock \bibinfo{title}{Heavy quark bound states in a quark-gluon plasma:
  {D}issociation and recombination},
\newblock \bibinfo{journal}{Nucl. Phys. A} \bibinfo{volume}{946}
  (\bibinfo{year}{2016}) \bibinfo{pages}{49--88}.
%Type = Article
\bibitem[{Du et~al.(2017)Du, He, and Rapp}]{rapp17a}
\bibinfo{author}{X.~Du}, \bibinfo{author}{M.~He}, \bibinfo{author}{R.~Rapp},
\newblock \bibinfo{title}{{Color screening and regeneration of bottomonia in
  high-energy heavy-ion collisions}},
\newblock \bibinfo{journal}{Phys. Rev. C} \bibinfo{volume}{96}
  (\bibinfo{year}{2017}) \bibinfo{pages}{054901}.
%Type = Article
\bibitem[{Ferreiro and Lansberg(2018)}]{fela18}
\bibinfo{author}{E.~G. Ferreiro}, \bibinfo{author}{J.~P. Lansberg},
\newblock \bibinfo{title}{Is bottomonium suppression in proton-nucleus and
  nucleus-nucleus collisions at {LHC} energies due to the same effects?},
\newblock \bibinfo{journal}{JHEP} \bibinfo{volume}{2018} (\bibinfo{year}{2018})
  \bibinfo{pages}{094}.
%Type = Article
\bibitem[{Boyd et~al.(2019)Boyd, Cook, Islam, and Strickland}]{strickland-2019}
\bibinfo{author}{J.~Boyd}, \bibinfo{author}{T.~Cook},
  \bibinfo{author}{A.~Islam}, \bibinfo{author}{M.~Strickland},
\newblock \bibinfo{title}{{Heavy quarkonium suppression beyond the adiabatic
  limit}},
\newblock \bibinfo{journal}{Phys. Rev. D} \bibinfo{volume}{100}
  (\bibinfo{year}{2019}) \bibinfo{pages}{076019}.
%Type = Article
\bibitem[{Rothkopf(2020)}]{rot20}
\bibinfo{author}{A.~Rothkopf},
\newblock \bibinfo{title}{Heavy quarkonium in extreme conditions},
\newblock \bibinfo{journal}{Phys. Rep.} \bibinfo{volume}{858}
  (\bibinfo{year}{2020}) \bibinfo{pages}{1--117}.
%Type = Article
\bibitem[{Brambilla et~al.(2023)Brambilla, Escobedo, Islam, Strickland, Tiwari,
  Vairo, and Griend}]{bram23}
\bibinfo{author}{N.~Brambilla}, \bibinfo{author}{M.~A. Escobedo},
  \bibinfo{author}{A.~Islam}, \bibinfo{author}{M.~Strickland},
  \bibinfo{author}{A.~Tiwari}, \bibinfo{author}{A.~Vairo},
  \bibinfo{author}{P.~V. Griend},
\newblock \bibinfo{title}{Regeneration of bottomonia in an open quantum system
  approach},
\newblock \bibinfo{journal}{Phys. Rev. D} \bibinfo{volume}{946}
  (\bibinfo{year}{2023}) \bibinfo{pages}{L011502}.
%Type = Article
\bibitem[{Nendzig and Wolschin(2014)}]{ngw14}
\bibinfo{author}{F.~Nendzig}, \bibinfo{author}{G.~Wolschin},
\newblock \bibinfo{title}{Bottomium suppression in {PbPb} collisions at {LHC}
  energies},
\newblock \bibinfo{journal}{J. Phys. G: Nuclear and Particle Physics}
  \bibinfo{volume}{41} (\bibinfo{year}{2014}) \bibinfo{pages}{095003}.
%Type = Article
\bibitem[{Dinh et~al.(2019)Dinh, Hoelck, and Wolschin}]{dhw19}
\bibinfo{author}{V.~H. Dinh}, \bibinfo{author}{J.~Hoelck},
  \bibinfo{author}{G.~Wolschin},
\newblock \bibinfo{title}{Hot-medium effects on {{$\Upsilon$}} yields in {pPb}
  collisions at $\sqrt{s_{NN}}=8.16$ {TeV}},
\newblock \bibinfo{journal}{Phys. Rev. C} \bibinfo{volume}{100}
  (\bibinfo{year}{2019}) \bibinfo{pages}{024906}.
%Type = Article
\bibitem[{Braaten and Pisarski(1990)}]{brapi90}
\bibinfo{author}{E.~Braaten}, \bibinfo{author}{R.~Pisarski},
\newblock \bibinfo{title}{Resummation and gauge invariance of the gluon damping
  rate in hot {QCD}},
\newblock \bibinfo{journal}{Phys. Rev. Lett.} \bibinfo{volume}{64}
  (\bibinfo{year}{1990}) \bibinfo{pages}{1338--1341}.
%Type = Article
\bibitem[{Rebhan(1993)}]{reb93}
\bibinfo{author}{A.~Rebhan},
\newblock \bibinfo{title}{Non-abelian {Debye} mass at next-to-leading order},
\newblock \bibinfo{journal}{Phys. Rev. D} \bibinfo{volume}{48}
  (\bibinfo{year}{1993}) \bibinfo{pages}{R3967--R3970}.
%Type = Article
\bibitem[{{Laine} et~al.(2007){Laine}, {Philipsen}, {Tassler}, and
  {Romatschke}}]{laine07}
\bibinfo{author}{M.~{Laine}}, \bibinfo{author}{O.~{Philipsen}},
  \bibinfo{author}{M.~{Tassler}}, \bibinfo{author}{P.~{Romatschke}},
\newblock \bibinfo{title}{{Real-time static potential in hot QCD}},
\newblock \bibinfo{journal}{JHEP} \bibinfo{volume}{2007} (\bibinfo{year}{2007})
  \bibinfo{pages}{54}.
%Type = Article
\bibitem[{{Jacobs} et~al.(1986){Jacobs}, {Olsson}, and {Suchyta}}]{ja86}
\bibinfo{author}{S.~{Jacobs}}, \bibinfo{author}{M.~G. {Olsson}},
  \bibinfo{author}{C.~{Suchyta}, III},
\newblock \bibinfo{title}{{Comparing the Schr{\"o}dinger and spinless Salpeter
  equations for heavy-quark bound states}},
\newblock \bibinfo{journal}{Phys. Rev. D} \bibinfo{volume}{33}
  (\bibinfo{year}{1986}) \bibinfo{pages}{3338--3348}.
%Type = Article
\bibitem[{{Brambilla} et~al.(2008){Brambilla}, {Ghiglieri}, {Vairo}, and
  {Petreczky}}]{brambilla-etal-2008}
\bibinfo{author}{N.~{Brambilla}}, \bibinfo{author}{J.~{Ghiglieri}},
  \bibinfo{author}{A.~{Vairo}}, \bibinfo{author}{P.~{Petreczky}},
\newblock \bibinfo{title}{{Static quark-antiquark pairs at finite
  temperature}},
\newblock \bibinfo{journal}{Phys. Rev. D} \bibinfo{volume}{78}
  (\bibinfo{year}{2008}) \bibinfo{pages}{014017}.
%Type = Article
\bibitem[{{Beraudo} et~al.(2008){Beraudo}, {Blaizot}, and
  {Ratti}}]{beraudo-etal-2008}
\bibinfo{author}{A.~{Beraudo}}, \bibinfo{author}{J.~P. {Blaizot}},
  \bibinfo{author}{C.~{Ratti}},
\newblock \bibinfo{title}{{Real and imaginary-time Q{\= Q} correlators in a
  thermal medium}},
\newblock \bibinfo{journal}{Nucl. Phys. A} \bibinfo{volume}{806}
  (\bibinfo{year}{2008}) \bibinfo{pages}{312--338}.
%Type = Article
\bibitem[{{Karsch} et~al.(1988){Karsch}, {Mehr}, and {Satz}}]{karsch-etal-1988}
\bibinfo{author}{F.~{Karsch}}, \bibinfo{author}{M.~T. {Mehr}},
  \bibinfo{author}{H.~{Satz}},
\newblock \bibinfo{title}{{Color screening and deconfinement for bound states
  of heavy quarks}},
\newblock \bibinfo{journal}{Z. Phys. C} \bibinfo{volume}{37}
  (\bibinfo{year}{1988}) \bibinfo{pages}{617--622}.
%Type = Article
\bibitem[{{Bethke}(2013)}]{bethke-2013}
\bibinfo{author}{S.~{Bethke}},
\newblock \bibinfo{title}{{World Summary of {$\alpha$}$_{S}$ (2012)}},
\newblock \bibinfo{journal}{Nucl. Phys. B (Proc. Supp.)} \bibinfo{volume}{234}
  (\bibinfo{year}{2013}) \bibinfo{pages}{229--234}.
%Type = Article
\bibitem[{Brambilla et~al.(2005)Brambilla, Pineda, Soto, and Vairo}]{bram08}
\bibinfo{author}{N.~Brambilla}, \bibinfo{author}{A.~Pineda},
  \bibinfo{author}{J.~Soto}, \bibinfo{author}{A.~Vairo},
\newblock \bibinfo{title}{{Effective-field theories for heavy quarkonium}},
\newblock \bibinfo{journal}{Rev. Mod. Phys.} \bibinfo{volume}{77}
  (\bibinfo{year}{2005}) \bibinfo{pages}{1423}.
%Type = Article
\bibitem[{{Brezinski} and {Wolschin}(2012)}]{bgw12}
\bibinfo{author}{F.~{Brezinski}}, \bibinfo{author}{G.~{Wolschin}},
\newblock \bibinfo{title}{{Gluodissociation and screening of {$\Upsilon$}
  states in PbPb collisions at $\sqrt{s_{NN}} = 2.76 \; \rm TeV$}},
\newblock \bibinfo{journal}{Phys. Lett. B} \bibinfo{volume}{707}
  (\bibinfo{year}{2012}) \bibinfo{pages}{534--538}.
%Type = Article
\bibitem[{{Escobedo} et~al.(2013){Escobedo}, {Giannuzzi}, {Mannarelli}, and
  {Soto}}]{esco13}
\bibinfo{author}{M.~A. {Escobedo}}, \bibinfo{author}{F.~{Giannuzzi}},
  \bibinfo{author}{M.~{Mannarelli}}, \bibinfo{author}{J.~{Soto}},
\newblock \bibinfo{title}{{Heavy quarkonium moving in a quark-gluon plasma}},
\newblock \bibinfo{journal}{Phys. Rev. D} \bibinfo{volume}{87}
  (\bibinfo{year}{2013}) \bibinfo{pages}{114005}.
%Type = Article
\bibitem[{{Vaccaro} et~al.(2013){Vaccaro}, {Nendzig}, and
  {Wolschin}}]{vaccaro-etal-2013}
\bibinfo{author}{F.~{Vaccaro}}, \bibinfo{author}{F.~{Nendzig}},
  \bibinfo{author}{G.~{Wolschin}},
\newblock \bibinfo{title}{The influence of the $\chi_{b}(3{P})$ state on the
  decay cascade of bottomium in {PbPb} collisions at {LHC} energies},
\newblock \bibinfo{journal}{EPL} \bibinfo{volume}{102} (\bibinfo{year}{2013})
  \bibinfo{pages}{42001}.
%Type = Article
\bibitem[{Navas et~al.(2024)}]{pdg24}
\bibinfo{author}{S.~Navas}, et~al. (\bibinfo{collaboration}{Particle Data
  Group}),
\newblock \bibinfo{title}{{Review of Particle Physics}},
\newblock \bibinfo{journal}{Phys. Rev. D} \bibinfo{volume}{110}
  (\bibinfo{year}{2024}) \bibinfo{pages}{030001}.
%Type = Article
\bibitem[{{Daghighian} and {Silverman}(1987)}]{daghighian-silverman-1987}
\bibinfo{author}{F.~{Daghighian}}, \bibinfo{author}{D.~{Silverman}},
\newblock \bibinfo{title}{{Relativistic formulation of the radiative
  transitions of charmonium and b-quarkonium}},
\newblock \bibinfo{journal}{Phys. Rev. D} \bibinfo{volume}{36}
  (\bibinfo{year}{1987}) \bibinfo{pages}{3401--3416}.
%Type = Article
\bibitem[{Bjorken(1983)}]{bjorken-1983}
\bibinfo{author}{J.~D. Bjorken},
\newblock \bibinfo{title}{{Highly relativistic nucleus-nucleus collisions: The
  central rapidity region}},
\newblock \bibinfo{journal}{Phys. Rev. D} \bibinfo{volume}{27}
  (\bibinfo{year}{1983}) \bibinfo{pages}{140--151}.
%Type = Article
\bibitem[{{Baym} et~al.(1983){Baym}, {Friman}, {Blaizot}, {Soyeur}, and
  {Czy{\.z}}}]{baym-etal-1983}
\bibinfo{author}{G.~{Baym}}, \bibinfo{author}{B.~L. {Friman}},
  \bibinfo{author}{J.-P. {Blaizot}}, \bibinfo{author}{M.~{Soyeur}},
  \bibinfo{author}{W.~{Czy{\.z}}},
\newblock \bibinfo{title}{{Hydrodynamics of ultra-relativistic heavy ion
  collisions}},
\newblock \bibinfo{journal}{Nucl. Phys. A} \bibinfo{volume}{407}
  (\bibinfo{year}{1983}) \bibinfo{pages}{541--570}.
%Type = Article
\bibitem[{{Gyulassy} and {Matsui}(1984)}]{gyulassi-matsui-1984}
\bibinfo{author}{M.~{Gyulassy}}, \bibinfo{author}{T.~{Matsui}},
\newblock \bibinfo{title}{{Quark-gluon-plasma evolution in scaling
  hydrodynamics}},
\newblock \bibinfo{journal}{Phys. Rev. D} \bibinfo{volume}{29}
  (\bibinfo{year}{1984}) \bibinfo{pages}{419--425}.
%Type = Article
\bibitem[{Adam et~al.(2017)}]{alice17}
\bibinfo{author}{J.~Adam}, et~al. (\bibinfo{collaboration}{ALICE
  Collaboration}),
\newblock \bibinfo{title}{Centrality dependence of the pseudorapidity density
  distribution for charged particles in {Pb-Pb} collisions at
  $\sqrt{s_{NN}}=5.02$ {TeV}},
\newblock \bibinfo{journal}{Phys. Lett. B} \bibinfo{volume}{772}
  (\bibinfo{year}{2017}) \bibinfo{pages}{567--577}.
%Type = Article
\bibitem[{Press et~al.(1993)Press, Teukolsky, Vetterling, and Flannery}]{pre93}
\bibinfo{author}{W.~H. Press}, \bibinfo{author}{S.~A. Teukolsky},
  \bibinfo{author}{W.~T. Vetterling}, \bibinfo{author}{B.~P. Flannery},
\newblock \bibinfo{title}{Numerical recipes in {C}: The art of scientific
  computing},
\newblock \bibinfo{journal}{Cambridge University Press, 2nd edition}
  (\bibinfo{year}{1993}).
%Type = Article
\bibitem[{Aaij et~al.(2023)}]{lhcb23}
\bibinfo{author}{R.~Aaij}, et~al. (\bibinfo{collaboration}{LHCb
  Collaboration}),
\newblock \bibinfo{title}{Measurement of ${\Upsilon}$ production in pp
  collisions at $\sqrt{s} = 5 \; \rm {TeV}$},
\newblock \bibinfo{journal}{JHEP} \bibinfo{volume}{07} (\bibinfo{year}{2023})
  \bibinfo{pages}{069}.
%Type = Article
\bibitem[{Aad et~al.(2023)}]{atlas23}
\bibinfo{author}{G.~Aad}, et~al. (\bibinfo{collaboration}{ATLAS
  Collaboration}),
\newblock \bibinfo{title}{Production of ${\Upsilon}{(nS)}$ mesons in {Pb+Pb}
  and pp collisions at 5.02 {TeV}},
\newblock \bibinfo{journal}{Phys. Rev. C} \bibinfo{volume}{107}
  (\bibinfo{year}{2023}) \bibinfo{pages}{054912}.
%Type = Article
\bibitem[{Lohs et~al.(1974)Lohs, Wolschin, and H{\"u}fner}]{lwh74}
\bibinfo{author}{K.-P. Lohs}, \bibinfo{author}{G.~Wolschin},
  \bibinfo{author}{J.~H{\"u}fner},
\newblock \bibinfo{title}{Nuclear {Auger} effect in muonic atoms},
\newblock \bibinfo{journal}{Nucl. Phys. A} \bibinfo{volume}{236}
  (\bibinfo{year}{1974}) \bibinfo{pages}{457--468}.
%Type = Article
\bibitem[{Kharzeev and Thews(1999)}]{kt99}
\bibinfo{author}{D.~Kharzeev}, \bibinfo{author}{R.~L. Thews},
\newblock \bibinfo{title}{Quarkonium formation time in a model-independent
  approach},
\newblock \bibinfo{journal}{Phys. Rev. C} \bibinfo{volume}{60}
  (\bibinfo{year}{1999}) \bibinfo{pages}{041901}.

\end{thebibliography}
%\end{thebibliography}
%\input stopping_plb_22.bbl
\end{document}